\documentclass[%
aip,
jcp,%
amsmath,amssymb,
reprint%
,longbibliography
]{revtex4-2}

\usepackage{amsmath}
\usepackage{amsfonts}
\usepackage{amssymb,bm}
\usepackage{graphicx}
\usepackage{physics}
\usepackage{upgreek}
\usepackage[outercaption]{sidecap}   
\usepackage{color}
\usepackage{hyperref}
\hypersetup{colorlinks=true,
	linkcolor=magenta,
	allcolors=magenta}
\usepackage{newtxtext}
\usepackage{newtxmath}
\usepackage[version=4]{mhchem}

\DeclareMathAlphabet{\pazocal}{OMS}{zplm}{m}{n}

\newcommand{\el}{\ensuremath{\mathrm{e}}}

\newcommand{\sing}{\ensuremath{\mathrm{S}}}
\newcommand{\trip}{\ensuremath{\mathrm{T}}}

\renewcommand{\op}[1]{\ensuremath{\hat{#1}}}

\begin{document}
	\title{The origin of chirality induced spin selectivity in photo-induced electron transfer}
\author{Thomas P. Fay\looseness=-1}
\email{tom.patrick.fay@gmail.com\looseness=-1}
\affiliation{Department of Chemistry, University of California, Berkeley, CA 94720, USA\looseness=-1}
\author{David T. Limmer\looseness=-1}
\affiliation{Department of Chemistry, University of California, Berkeley, CA 94720, USA\looseness=-1}
\affiliation{Kavli Energy Nanoscience Institute at Berkeley, Berkeley, CA 94720, USA\looseness=-1}
\affiliation{Chemical Sciences Division, Lawrence Berkeley National Laboratory, Berkeley, CA 94720, USA\looseness=-1}
\affiliation{Materials Science Division, Lawrence Berkeley National Laboratory, Berkeley, CA 94720, USA\looseness=-1}

	
	\begin{abstract}
		Here we propose a mechanism by which spin polarization can be generated dynamically in chiral molecular systems undergoing photo-induced electron transfer. The proposed mechanism explains how spin polarization emerges in systems where charge transport is dominated by incoherent hopping, mediated by spin orbit and electronic exchange couplings through an intermediate charge transfer state. We derive a simple expression for the spin polarization that predicts a non-monotonic temperature dependence consistent with recent experiments. We validate this theory using approximate quantum master equations and the numerically exact hierarchical equations of motion. The proposed mechanism of chirality induced spin selectivity should apply to many chiral systems, and the ideas presented here have implications for the study of spin transport at temperatures relevant to biology, and provide simple principles for the molecular control of spins in fluctuating environments.
	\end{abstract}

\maketitle


Chirality induced spin selectivity (CISS), in which molecular chirality controls spin polarization of electrons, has been observed in a wide variety of systems,\cite{Naaman2015,Naaman2020a,Waldeck2021} including in photo-induced electron transfer in artificial systems,\cite{Abendroth2019} such as quantum dots,\cite{Bloom2017} and biological molecules, such as photosystem I\cite{Carmeli2014} and DNA.\cite{SenthilKumar2013,Abendroth2017} Based on this, it has been suggested that CISS may serve some biological functions\cite{Michaeli2016} and that it could be exploited in nano- and molecular-scale spintronics technologies.\cite{Naaman2015,Brandt2017} Here we focus on the CISS effect in photo-induced molecular electron transfer.\cite{Carmeli2014} This effect is particularly puzzling, as it goes against the conventional notion that the initial singlet spin state of the system is preserved in this process.\cite{Wasielewski2006}

Theoretical descriptions of CISS have primarily been confined to either the coherent regime,\cite{Dalum2019,Hu2020,Geyer2020,Gutierrez2012,Michaeli2019} where electron transport is treated as a coherent tunneling process through a chiral molecular junction between leads, or static spin polarizations produced by spin-orbit coupling in the ground state of chiral molecules.\cite{Dianat2020,Fransson2020,Fransson2021} However photo-induced electron transfer in molecules like photosystem I, where electrons incoherently hop between a discrete set of donor and acceptor sites, cannot be described by either of these theories.\cite{Muller2010} In such systems coupling of the electron transfers to molecular vibrations is strong,\cite{Blumberger2015} and electronic exchange interactions can also be significant.\cite{Wasielewski2006} This necessitates the development of a theory of CISS that is appropriate for this incoherent regime of electron transport. Fortunately such techniques have been developed, primarily for understanding rates and dynamics of electron transfer reactions,\cite{May2000,Nitzan2006} and through the tools of quantum master equations such theories can be adapted to describe spin transport.\cite{Fay2018,Fay2021a} These techniques have recently been used to show that spin coherence, but no spin polarization, arises in a single electron transfer step in a chiral environment mediated by spin-orbit coupling.\cite{Fay2021a} Here we expand on this theory, and show how the interplay of chirality induced spin coherence, electron hopping, and electronic exchange coupling can dynamically produce large spin polarizations in electron transfer reactions. The theory presented here will be shown to explain the observed temperature dependence of CISS in photosystem I, and it also provides a framework for engineering molecules to maximize the CISS effect.

Let us start by reviewing how photo-initiated charge transfer, between a donor \ce{D} and acceptor \ce{A}, in chiral molecules generates spin coherence in charge transfer states, as described in Ref.~\onlinecite{Fay2021a}. This process generally proceeds via an electron transfer reaction from a bright locally excited singlet state,  \ce{S_1} = \ce{D^*-A}, to a charge transfer state, \ce{CT} = \ce{D^{$\bullet$+}-A^{$\bullet$-}}, which can exist in either a singlet or triplet spin state,
\begin{align*}
	\ce{S_1 -> CT}.
\end{align*}
As discussed in Ref.~\onlinecite{Fay2021a} direct diabatic coupling generates the \ce{CT} state in a singlet $\ket{\sing} = (\ket{\uparrow_\mathrm{D}\downarrow_\mathrm{A}}-\ket{\downarrow_\mathrm{D}\uparrow_\mathrm{A}})/\sqrt{2}$ spin state, whereas spin-orbit coupling (SOC) generates the CT state in the $i\ket{\trip_0}=i(\ket{\uparrow_\mathrm{D}\downarrow_\mathrm{A}}+\ket{\downarrow_\mathrm{D}\uparrow_\mathrm{A}})/\sqrt{2}$ (taking the $z$ axis to be defined by the spin-orbit coupling vector).\cite{Fedorov2003} So overall this electron transfer, mediated by spin-orbit coupling in a chiral molecule, generates the CT state in a coherent superposition of singlet and $\ket{\trip_0}$ triplet spin states, and its initial spin density operator, $\op{\sigma}_{\ce{CT}}(t)$, is given by
\begin{align}
	\op{\sigma}_{\mathrm{CT}}(0) = \dyad{\psi_\theta},
\end{align}
where $\ket{\psi_\theta} = \cos\theta\ket{\sing} + i \sin\theta\ket{\trip_{0}}$. The mixing angle $\theta$, is determined by the relative strength of the spin-preserving diabatic coupling $V_{\mathrm{DA}}$ and the spin-orbit coupling $\Lambda_{\mathrm{DA}}$, as $\theta = \arctan(\Lambda_{\mathrm{DA}}/(2V_{\mathrm{DA}}))$, and because the sign of $\Lambda_{\mathrm{DA}}$ depends on the chirality of the molecule, the sign of $\theta$ also depends on chirality. The spin polarization of the CT state is defined as the difference in the $z$ component of the spins between D and A. This is given by the expectation value of $\Delta\op{P}_z = \op{S}_{\mathrm{D}z}-\op{S}_{\mathrm{A}z} = \dyad{\sing}{\trip_0} + \dyad{\trip_0}{\sing}$, so the spin polarization is proportional to the \textit{real} part of the singlet-triplet coherence, $\mel*{\sing}{\op{\sigma}_{\ce{CT}}}{\trip_0}$. However the CT state is generated with purely \textit{imaginary} singlet-triplet coherence, so no spin polarization is formed directly by the electron transfer.\cite{Fay2021a} 

Now let us consider how the spin state $\op{\sigma}_{\mathrm{CT}}$ evolves when there exists an exchange coupling between the donor and acceptor electron spins in the CT state. In this case the spin Hamiltonian for the CT state can be taken to be $\op{H}_\mathrm{CT} = - 2J \op{P}_\trip $ ($\op{P}_\trip$ is a triplet spin projection operator),\cite{Steiner1989} and the time-evolution of the spin state is given by
\begin{align}
	e^{- i \op{H}_\mathrm{CT} t /\hbar} \ket{\psi_\theta} = \cos\theta\ket{\sing} + i \sin\theta e^{2i J t/\hbar}\ket{\trip_{0}}.
\end{align}
We see that exchange coupling generates a complex-valued oscillating singlet-triplet coherence in the CT spin density operator, and thus the CT state now has an oscillating real part of its singlet-triplet coherence. This means that the exchange coupling produces an oscillating spin polarization in the CT state which is given by,
\begin{align}
	\ev{\Delta P_z(t)} = -\sin(2\theta) \sin(\frac{2Jt}{\hbar}).
\end{align}

\begin{figure}[t]
	\includegraphics[width=0.45\textwidth]{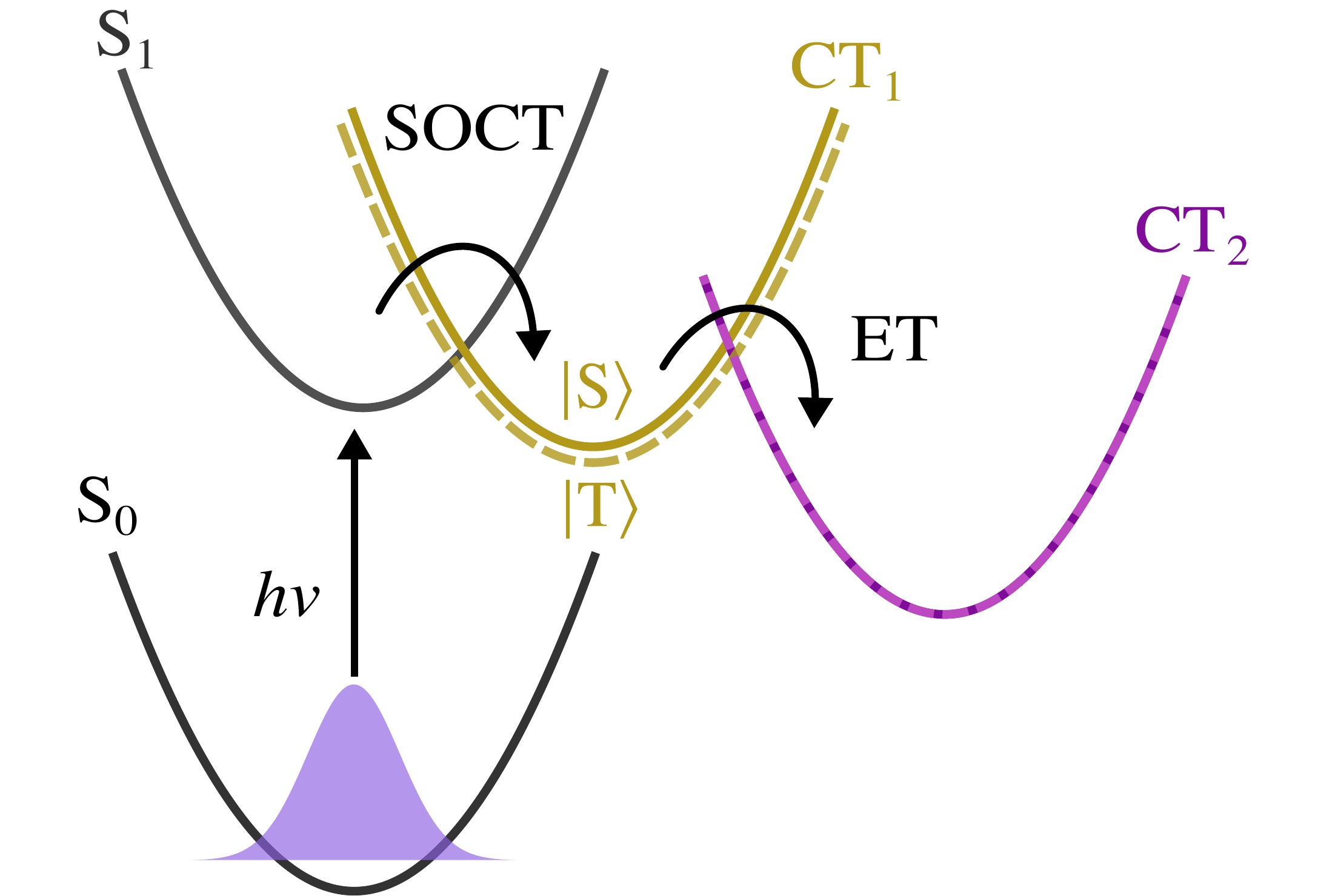}
	\caption{Schematic of the photo-induced charge transfer mechanism by which spin polarization can be generated. $\ce{S_1} = \ce{D^*-A_1-A_2}$, $\ce{CT_1} = \ce{D^{\bullet +}-A^{\bullet -}_1-A_2}$ and $\ce{CT_2} = \ce{D^{\bullet +}-A_1-A^{\bullet -}}_2$. Singlet states are shown as solid lines and triplet states as dashed lines. Note the non-negligible S-T gap in \ce{CT_1}, and the negligible S-T gap in \ce{CT_2}.}\label{hopping-model-fig}
\end{figure}
Overall the presence of exchange coupling effects a chirality-dependent transient oscillating spin polarization in the CT state. However this spin polarization clearly averages over time to zero. This naturally raises the question of how a \textit{static} spin polarization can be generated in a photo-induced charge transfer reaction. The answer lies in the fact that many photo-generated CT states are not formed by a simple direct electron transfer, but rather the final CT state is often formed via a sequence of downhill electron transfers via intermediate CT states.\cite{Wasielewski2006,Blumberger2015,Muller2010} The final CT state is often a well-separated radical ion pair state. In this state the exchange coupling between donor and acceptor spins is very weak, but the intermediate CT states often have a significantly larger exchange coupling, due to the closer proximity of the unpaired electrons.\cite{Wasielewski2006} In this way, imaginary-valued singlet-triplet coherence created through spin-orbit coupled charge transfer can be transiently converted into a net spin polarization between the donor and acceptor electrons in an intermediate CT state with a large exchange coupling. This spin polarization can be transferred to the final CT state via a subsequent incoherent spin state preserving electron transfer. The spin polarization transferred to the final CT no longer oscillates due to the weak exchange coupling in this state, thus a net static spin polarization is generated. This is the basic mechanism by which static spin polarization can be generated in multi-step electron transfer reactions in chiral molecules.

We will now explore this further, by considering the simplest multi-state model including only one intermediate CT state in the chiral system (i.e. a three state model),
\begin{align*}
	\ce{S_1 ->[$k_{\mathrm{ET}1}$]  CT_1 ->[$k_{\mathrm{ET}2}$]  CT_2}.
\end{align*}
We take the precursor state to be an excited donor state, $\ce{S_1} = \ce{D^*-A_1-A_2}$, the intermediate charge transfer state corresponds to an electron transfer from D to a primary acceptor \ce{A_1}, so $\ce{CT_1} = \ce{D^{\bullet +}-A^{\bullet -}_1-A_2}$, and the final charge transfer state \ce{CT_2} is taken as the state formed by electron transfer from the primary acceptor \ce{A_1} to the secondary acceptor \ce{A_2}, so $\ce{CT_2} = \ce{D^{\bullet +}-A_1-A^{\bullet -}}_2$. The CT states can exist in either singlet or triplet spin states but the \ce{S_1} state only exists in the singlet spin state. A schematic of this system is shown in Fig.~\ref{hopping-model-fig}. To start with, for simplicity we will assume these downhill electron transfers occur irreversibly, and we will also assume that only the first electron transfer is spin-dependent, and occurs via chiral spin-orbit coupled charge transfer. 

In this model, the spin density operators for each state obey the following set of equations,\cite{Fay2018,Fay2021a}
\begin{subequations}\label{qmes-eq}
\begin{align}
	\dv{t}\op{\sigma}_{\ce{S_1}}(t) &= - k_\mathrm{ET1} \op{\sigma}_{\ce{S_1}}(t) \\
	\begin{split}
		\dv{t}\op{\sigma}_{\ce{CT_1}}(t) &= -\frac{i}{\hbar}[\op{H}_{\ce{CT_1}}+{\delta\epsilon} \dyad{\psi_\theta},\op{\sigma}_{\ce{CT_1}}(t)] \\
		+k_\mathrm{ET1} &\dyad{\psi_\theta}{\sing}\op{\sigma}_{\ce{S_1}}(t)\dyad{\sing}{\psi_\theta} -k_\mathrm{ET2} \op{\sigma}_{\ce{CT_1}}(t) 
	\end{split}\\
	\begin{split}
		\dv{t}\op{\sigma}_{\ce{CT_2}}(t) &= -\frac{i}{\hbar}[\op{H}_{\ce{CT_2}},\op{\sigma}_{\ce{CT_2}}(t)] +k_\mathrm{ET2} \op{\sigma}_{\ce{CT_1}}(t) 
	\end{split}
\end{align}
\end{subequations}
Here $\op{H}_{\ce{CT}_1}$ and $\op{H}_{\ce{CT}_2}$ are the spin Hamiltonians for the $\ce{CT_1}$ and $\ce{CT_2}$ states, describing all the spin interactions in these states. Here the spin Hamiltonians are simplified to just include the exchange couplings, so $\op{H}_{\ce{CT_1}}=-2J\op{P}_{\trip}$ and $\op{H}_{\ce{CT_2}}=0$ (the energy differences between different CT states enter into the master equations via the rate constants). This set of coupled equations can be obtained straightforwardly using perturbative Nakajima-Zwanzig theory,\cite{Nakajima1958,Zwanzig1960} as has been described previously (a summary of the approximations in Refs.~\onlinecite{Fay2018} and \onlinecite{Fay2021a} will be discussed shortly). 
The shift term $\delta \epsilon$ naturally emerges from the master equation theory, and its value will depend on the details of the electron transfer process. For a large downhill driving force, $\Delta G_{\mathrm{ET1}}$, it can be estimated as $\delta \epsilon \approx -(V_{\mathrm{DA_1}}^2 + (\Lambda_{\mathrm{DA_1}}/2)^2)/(\lambda_{\mathrm{ET1}}-\Delta G_{\mathrm{ET1}})$, where $\lambda_{\mathrm{ET1}}$ is the reorganization energy for the first electron transfer. We can also understand the $\delta\epsilon$ term as a net chiral superexchange spin-orbit coupling term, which couples $\sing$ and $\trip_0$ states via the \ce{S_1} state. 

One can then solve these Eq.~\eqref{qmes-eq} analytically for an initial state where only \ce{S_1} is populated, with $\op{\sigma}_{\ce{S_1}}(0) = \dyad{\sing}$, from which the final spin polarization in \ce{CT_2}, given by $\ev{\Delta P_z} =\lim_{t\to \infty} \tr[(\op{S}_{\mathrm{D}z}-\op{S}_{\mathrm{A}z})\op{\sigma}_{\ce{CT_2}}(t)]$, is found to be
\begin{align}\label{Pz-eq}
	\ev{\Delta P_z} = -\frac{k_{\mathrm{ET2}}(2J/\hbar) \sin(2\theta)}{ k_{\mathrm{ET2}}^2  + \Omega^2},
\end{align}
where $\hbar \Omega = \sqrt{4J^2 + {\delta\epsilon}^2 +4J {\delta\epsilon} \cos(2\theta)}$. Simultaneously a zero quantum coherence (which depends on the imaginary part of the $\sing$-$\trip_0$ coherence) is produced 
\begin{align}
	\ev{S_{\mathrm{D}x}S_{\mathrm{A}y}\!-\!S_{\mathrm{D}y}S_{\mathrm{A}x}} \!=\! \sin(2\theta) \!\left(1\! -  \frac{2(J/\hbar)^2}{k_{\mathrm{ET2}}^2 + \Omega^2} \!-\! \frac{J {\delta\epsilon} \cos(2\theta)}{(\hbar\Omega)^2}\!\right)\!.
\end{align}
This zero quantum coherence is produced directly by the initial SOC mediated electron transfer from \ce{S_1}, and modified by the spin dynamics in \ce{CT_1}.\cite{Fay2021a} Lastly, we find the final singlet population is given by
\begin{align}
	\ev{\rho_\sing} = \cos^2\theta \left(1 + \frac{2J\delta\epsilon / \hbar^2}{k_{\mathrm{ET2}}^2+\Omega^2} - \frac{2J\delta \epsilon \cos(2\theta)}{(\hbar\Omega)^2}\right)
\end{align}
and the $\trip_0$ population is given by $\ev*{\rho_{\trip_0}} = 1 - \ev*{\rho_\sing}$. We see the final spin density operator for the \ce{CT_2} is a combination of the phenomenological theory in Ref.~\onlinecite{Luo2021} and the theory obtained from the one-step transfer model in Ref.~\onlinecite{Fay2021a}. 

Let us now discuss a few points of interest on the expression for the spin polarization in Eq.~\eqref{Pz-eq}. Firstly, this expression is an odd function of $\theta = \arctan({\Lambda}_{\mathrm{DA_1}}/2V_{\mathrm{DA_1}})$, so on changing the chirality of the system the sign of this polarization is reversed, because the sign of $\Lambda_{\mathrm{DA_1}}$ changes, which confirms that this corresponds to a CISS effect. Secondly, in the absence of the shift term $\delta \epsilon$ the spin polarization is bounded by $|\ev*{\Delta P_z}| \leq |\sin(2\theta)|$ as $k_{\mathrm{ET1}}/(2J)$ is varied. In the limit of weak spin-orbit coupling, this bound is simply $|\Lambda_{\mathrm{DA_1}}/V_{\mathrm{DA_1}}|$, which is the ratio of the spin orbit coupling to the diabatic coupling for the first electron transfer. However when the shift term is included it is possible for $\Omega^2 < (2J)^2$. For example if $J$ and $\delta\epsilon$ have opposite signs and $\cos(2\theta)>0$, the polarization can be enhanced significantly beyond the maximum that one predicts in the absence of $\delta \epsilon$. 

\begin{figure*}[t]
	\centering
		\includegraphics[width=0.85\textwidth]{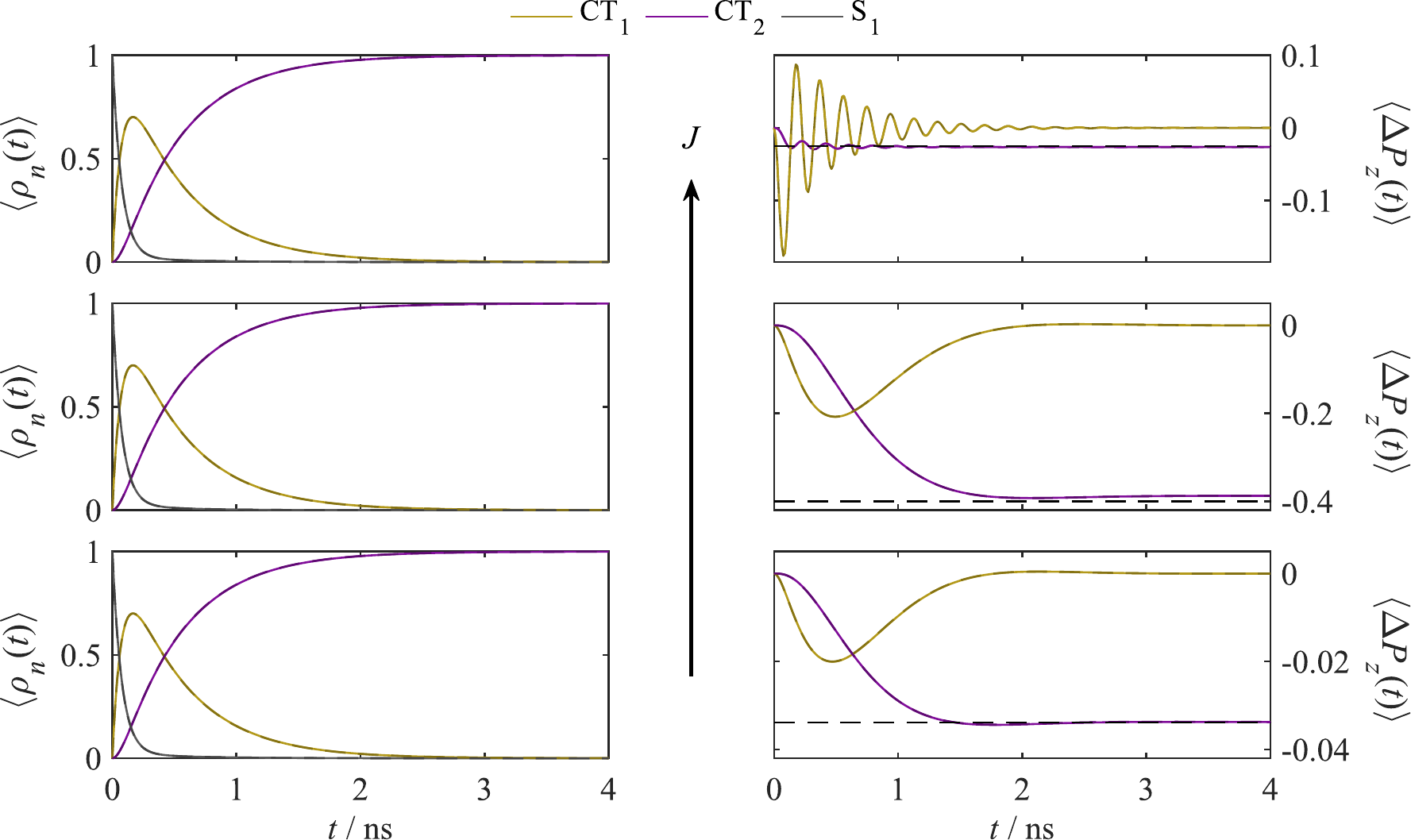}
		\caption{Population and spin polarization dynamics for a three CT state model of electron transfer calculated with HEOM (solid lines) and the second order QME (dashed lines). Left column: total populations of \ce{S_1} (grey), \ce{CT_1} (orange) and \ce{CT_2} (purple) as a function of time. Right column: spin polarization in \ce{CT_1} (orange) and \ce{CT_2} (purple) as a function of time, with the value predicted by Eq.~\eqref{Pz-eq} indicated by the black dashed line. From top to bottom the value of the exchange coupling in the \ce{CT_1} state is varied with  $J/g_\el\mu_\mathrm{B}=$ 100 mT (top), 10 mT (middle) and 1 mT (bottom), where $\mu_\mathrm{B}$ is the Bohr magneton and $g_\el$ is the $g$ factor for the free electron spin.}\label{qme-heom-fig}
\end{figure*}
As a final point, we note that the magnitude of the spin polarization $|\ev*{P_z}|$ goes through a maximum when $k_{\mathrm{ET2}} = \Omega$. This can be understood simply as follows. When $k_{\mathrm{ET2}}$ is very high compared to the frequency at which polarization in \ce{CT_1} oscillates, $\Omega$, then there is no time for spin polarization to build up in \ce{CT_1} before this state decays to \ce{CT_2}. Conversely when $k_\mathrm{ET2}$ is very low compared to $\Omega$, the spin polarization is averaged to zero by the oscillations as it is transferred, so no net spin polarization in \ce{CT_2} is observed. This observation has important implications for the temperature dependence of $\ev{\Delta P_z}$, as we will discuss shortly.

In order to further explore this theory, and to demonstrate the validity of the perturbative quantum master equations (QMEs) and additional approximations used to obtain the expressions above, we performed simulations of the three state condensed phase electron transfer process illustrated in Fig.~\ref{hopping-model-fig}. We employ both numerically exact quantum dynamics and the approximate QMEs, with the nuclear degrees of freedom treated as a harmonic bath. The total Hamiltonian in this model is a sum of nuclear (bath) kinetic energy, potential energy terms for the different states, and a diabatic/SOC term (further details are given in the SI, Eq.~(S.1)). Using the Hierarchical Equations of Motion (HEOM), we can compute the exact dynamics when the potentials are harmonic.\cite{Ishizaki2005}  

For comparison, we also perform simulations of this model with the full perturbative QMEs, Eq.~\eqref{qmes-eq}, modified to include back reaction terms.\cite{Fay2021a} To briefly summarise the approximations used in Refs.~\onlinecite{Fay2018} and \onlinecite{Fay2021a} to obtain these, in these QMEs we assume the coherences between different electron transfer states, e.g. $\dyad{\ce{S}_1}{\ce{CT_1},\sing}$, are short-lived and small (due to the large differences between potential energy surfaces), so these coherences are projected out using Nakajima-Zwanzig theory (but \textit{spin} coherences are retained). The Nakajima-Zwanzig kernel is then expanded at lowest order in the diabatic couplings, with spin Hamiltonian terms in the kernel ignored (due to the large separation in energy scale between the nuclear motion and spin dynamics), and the population transfer dynamics are treated as Markovian. For the harmonic bath model given above, the second order QME parameters (rate constants and $\delta \epsilon$) can be computed exactly from the spectral density,\cite{Fay2018} details of which are given in the SI. In the high temperature limit, these rates can be well approximated with Marcus theory, $k_{\mathrm{ET}} = (1/\hbar)\Gamma_{\mathrm{DA}}^2(\pi/\lambda_{\mathrm{ET}}k_\mathrm{B} T)^{1/2} e^{-(\Delta G_{\mathrm{ET}}+\lambda_{\mathrm{ET}})^2/4\lambda_{\mathrm{ET}} k_\mathrm{B}T}$, where $\Gamma_{\mathrm{DA}} = (V_\mathrm{DA}^2+\Lambda_{\mathrm{DA}}^2/4)^{1/2}$ is the combined diabatic/SOC coupling for the electron transfer.\cite{May2000} Although we only consider harmonic potential energy surfaces here, the QMEs apply to general potential energy surfaces for the electron transfer states.

We set the free energy changes of the electron transfers to be $\Delta G_{\mathrm{ET1}} = -0.1\text{ eV}$ and $\Delta G_{\mathrm{ET2}} = -0.25\text{ eV}$, with reorganisation energies of $\lambda_{\mathrm{ET1}}=0.1\text{ eV}$ and $\lambda_{\mathrm{ET2}}=0.2\text{ eV}$. Diabatic state couplings are taken as $\Gamma_{\mathrm{DA_1}} = (V_{\mathrm{DA}_1}^2+\Lambda_{\mathrm{DA_1}}^2/4)^{1/2}= 0.5\text{ meV}$ (\ce{S_1}-\ce{CT_1} coupling) and $\Gamma_{\mathrm{A_1 A_2}} = V_{\mathrm{A_1A_2}}=0.25\text{ meV}$ (\ce{CT_1}-\ce{CT_2} coupling), and $\theta = \pi/16$, which corresponds to $\sim\!4\%$ of \ce{CT_1} molecules being formed in the triplet state.\cite{Maeda2011a,Zhukov2020,Fay2019b} In these models we treat the intramolecular vibrational modes and solvent bath as a single harmonic bath with a Debye spectral density,\cite{Ishizaki2005} $\mathcal{J}(\omega) = 2\lambda_{\mathrm{D}} \gamma_\mathrm{D}\omega/(\gamma_\mathrm{D}^2+\omega^2)$,with a cut-off frequency of $\hbar \gamma_\mathrm{D} = 0.05 k_\mathrm{B} T$, at $T = 300\text{ K}$, and with $\lambda_{\mathrm{D}} = \lambda_{\mathrm{ET1}}$. These parameters are chosen to be representative of electron transfers in typical organic donor-acceptor systems,\cite{Blumberger2015,Makita2017} whilst being computationally tractable with the HEOM method. 

Figure \ref{qme-heom-fig} shows the results of these simulations for a range of values of the $J$ coupling (decreasing from top to bottom) in the intermediate state for populations (left) and spin polarizations (right). The population dynamics clearly show that hopping dominates the ET process in these examples, with a significant transient population of the \ce{CT_1} appearing on a time-scale of $\sim 0.25\text{ ns}$ and decaying on a time-scale of $\sim 1\text{ ns}$. We see the final spin polarization in \ce{CT_2} increases and then decreases with decreasing $J$ (note the different $\ev{\Delta P_z(t)}$ scales from top to bottom), as predicted by the simple theory in Eq.~\eqref{Pz-eq}. Large spin polarizations, on the order of 40\%, can be observed even with a small fraction of triplet states formed in ET1. The second order QMEs in this example are very accurate for both population and spin polarization dynamics, which validates the approximations of Markovian dynamics and spin-independent rate constants invoked to derive these. The prediction from Eq.~\eqref{Pz-eq} for the final spin polarization (shown as a black dashed line in the left panels of Fig.~\ref{qme-heom-fig}), provides an accurate estimate in these examples, with small deviations due to back reactions.


\begin{figure}[b]
	\includegraphics[width=0.4\textwidth]{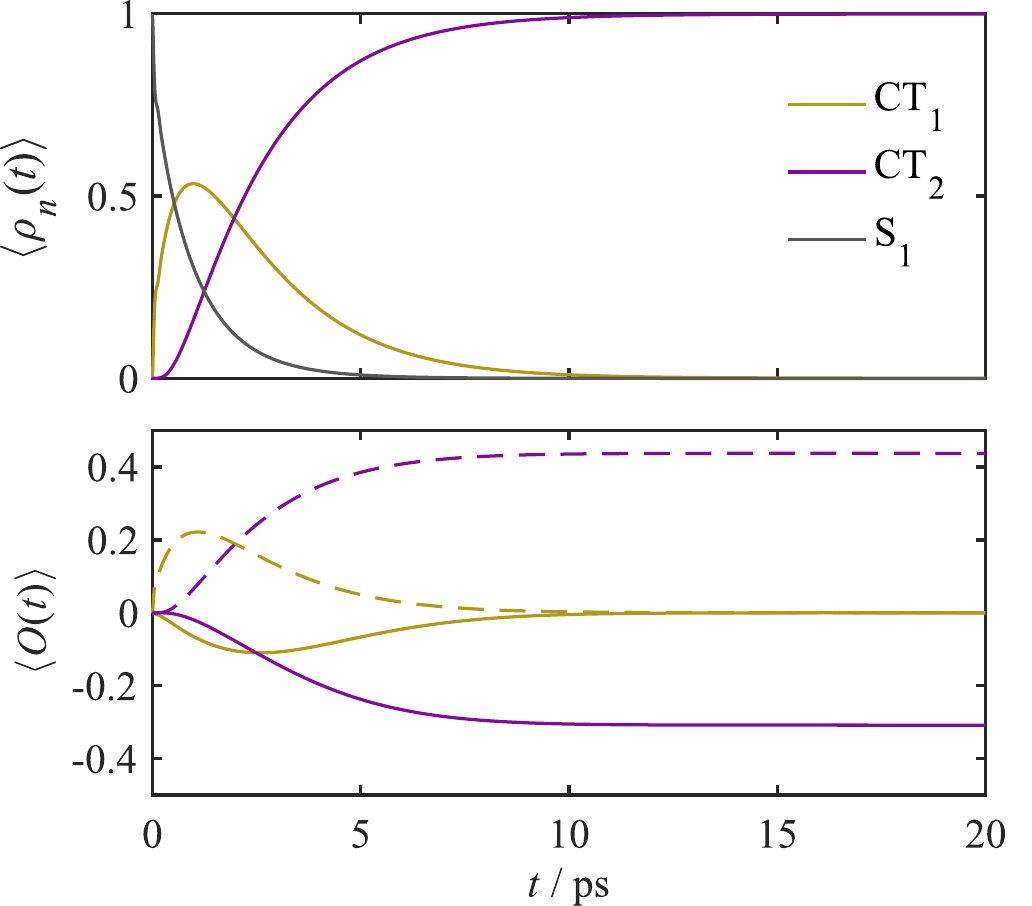}
	\caption{HEOM dynamics of the three state model with larger diabatic state couplings. Populations (top panel), and spin polarization $O= {\Delta P_z}$ and zero quantum coherences $O = {S_{\mathrm{D}x}S_{\mathrm{A}y}-S_{\mathrm{D}y}S_{\mathrm{A}x}}$ (bottom panel) with the same colour scheme as Fig.~\ref{qme-heom-fig}. Other ET parameters are the same as in Fig.~\ref{qme-heom-fig}.}\label{large-coup-fig}
\end{figure}
In the above examples we have only considered a regime where second order perturbation theory is valid for the rate constants and $\delta \epsilon$. The general form of the master equation presented above is only slightly altered on including higher order effects in the electronic coupling\cite{Fay2018,Fay2021a}; specifically we should add a decoherence term of the form $-(k_{\mathrm{D}}/2)[\dyad{\psi_\theta},[\dyad{\psi_\theta}, \op{\sigma}_{\ce{CT_1}}(t)]]$ to the equation for $\op{\sigma}_{\ce{CT_1}}(t)$ .\cite{Fay2021a} This causes the spin polarization in the intermediate state to decay, but for sufficiently large $2J$ significant spin polarization can still emerge and subsequently be transferred to the final CT state. This is demonstrated in Fig.~\ref{large-coup-fig} where we show the population dynamics (top panel), spin polarization (bottom panel, solid lines), and zero quantum coherence (bottom panel, dashed lines) for the three state model with $\Gamma_{\mathrm{DA_1}}=10\text{ meV}$, $V_{\mathrm{A_1A_2}}=5\text{ meV}$ and $J/g_e\mu_\mathrm{B} =2\text{ T}$. In this limit the second order rate constant for the first electron transfers are approximately a factor of 5 too large compared to those obtained by fitting the population dynamics, so this example is clearly outside of the limits of second order perturbation theory. However, we still clearly see the emergence of a large static spin polarization ($\sim 30\%$) in the \ce{CT_2} state, as well as significant zero quantum coherence. This demonstrates that this mechanism can produce significant spin polarizations even outside the second order limit.

\begin{figure}[t]
	\includegraphics[width=0.475\textwidth]{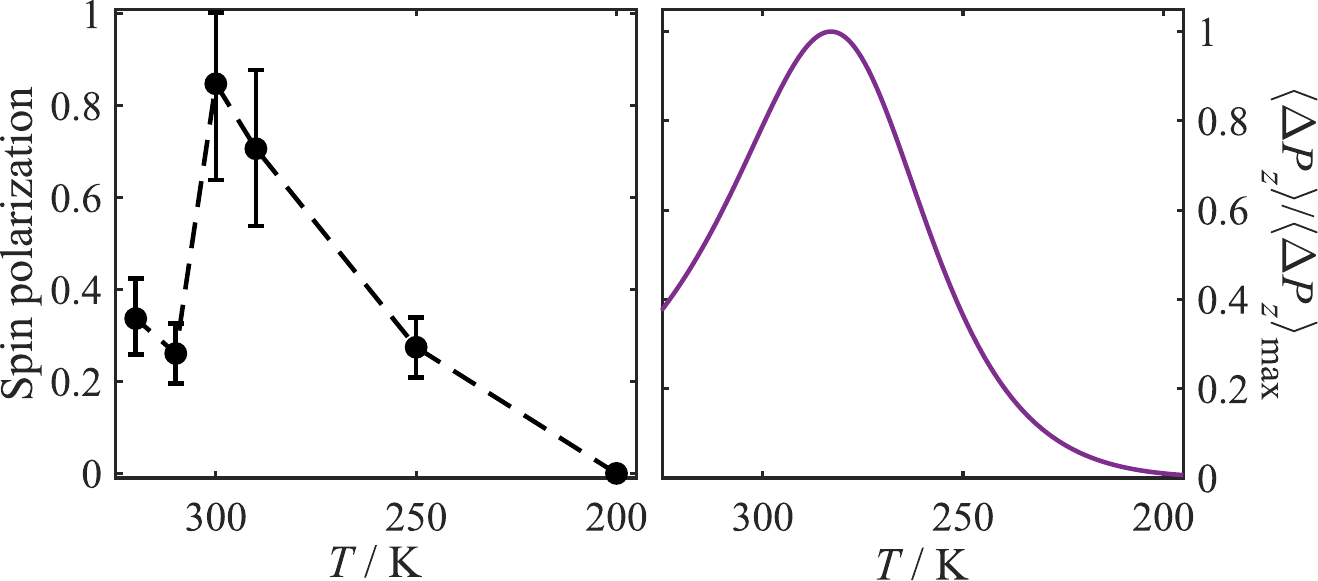}
	\caption{Experimental data from Ref.~\onlinecite{Carmeli2014} for the corrected spin-polarization in photosystem I (left), and a model based on Eq.~\eqref{Pz-eq} for the spin polarization relative to its maximum value (right), $\ev{\Delta P_z}/\ev{\Delta P_z}_{\mathrm{max}}= 2 aT^{-1/2}e^{-E_a/k_\mathrm{B}T}[1+(aT^{-1/2}e^{-E_a/k_\mathrm{B}T})^2]^{-1}$. The fitted parameters are $a = 8.346\times 10^6\text{ K}^{1/2}$ and $E_a = 0.32\text{ eV}$.}\label{carmeli-fig}
\end{figure}
Having validated the general theory on a model condensed phase electron transfer, let us consider how this hopping mechanism predicts a temperature dependent spin polarization in chiral systems. Returning to Eq.~\eqref{Pz-eq} for $\ev{\Delta P_z}$, we can see how this  temperature dependence emerges. Typically the exchange coupling, and energy shift terms are, to a good approximation, independent of temperature,\cite{Fay2018} whereas if the electron transfer is activated, $k_{\mathrm{ET2}}$ will have an exponential dependence on inverse temperature.\cite{Blumberger2015} Therefore increasing $T$ will generally increase the ratio $k_{\mathrm{ET2}}/\Omega$, so the spin polarization will go through a maximum at a particular temperature, where $k_{\mathrm{ET2}}/\Omega = 1$. Interestingly this temperature dependence of the spin polarization has been observed in the experiments of Carmeli \textit{et al.} on CISS in electron transfer in photosystem I,\cite{Carmeli2014} which is known to proceed via several electron hopping steps.\cite{Muller2010} We have fitted the experimental temperature dependent spin polarization, relative to its maximum value, for photosystem I to a model based on Eq.~\eqref{Pz-eq} assuming Marcus theory\cite{Marcus1956} can be applied to describe the temperature dependence of the rate constants. The experimental data and model are shown in Fig.~\ref{carmeli-fig}, where we see this model is consistent with the available experimental data, and the fitted activation energy of $E_a = 0.32\text{ eV}$ is in line with the activation energies of the electron transfers in photosystem I.\cite{Makita2017} 
It should be noted that the spin-polarization reported in these experiments exceeds the theoretical maximum predicted by Eq.~\eqref{Pz-eq} of 50\%. We believe this is due to the error in the correction procedure employed in Ref.~\onlinecite{Carmeli2014} for obtaining the spin polarizations, which, as discussed by the authors, introduces a large error in the corrected spin polarization. The ``uncorrected'' experimental spin polarizations, which have a maximum value of <10\%, still have a non-monotonic dependence on temperature,\cite{Carmeli2014} which is still consistent with our proposed mechanism of CISS. 

To conclude, in this work we have shown how photo-induced spin-orbit coupled charge transport in chiral systems, together with multistep charge transport via CT states with significant exchange couplings, generates the CISS spin polarization effect. We also see that a zero quantum coherence in the final CT state emerges. The CISS effect described here depends strongly on electron transfer parameters, in particular the rate constant for downhill electron transfer for intermediate CT states. This predicts a specific form for the temperature dependence of the spin polarization, which is consistent with the experiments of Carmeli \textit{et al.} in Ref.~\onlinecite{Carmeli2014}. 

This theory of the CISS effect in molecular systems opens the door to many potential avenues of research. With a theory for the ZQCs and spin polarizations generated in chiral molecules, well-established principles for synthetically tuning electronic properties of donor-acceptor systems\cite{Wasielewski2006,Wasielewski2020} could be applied to engineer systems displaying large CISS effects, which can be probed using a variety of possible experiments.\cite{Abendroth2019,Waldeck2021,Fay2021a,Luo2021,Chiesa2021} In fact the first examples of such molecules have already been synthesised.\cite{Junge2020} These systems could find an array of uses in single molecule-scale quantum computing and quantum information applications.\cite{Wasielewski2020,Naaman2015} It is also known that electron hopping occurs in many important biological electron transfer systems,\cite{Blumberger2015} such as in photosynthesis in photosystem I\cite{Muller2010} and in cryptochromes, which are involved in signalling and potentially magnetoreception.\cite{Dodson2013} What role role could molecular CISS play in these systems?\cite{Michaeli2016} For example, could biological systems have evolved to inhibit singlet-selective electron recombination reactions using CISS? Equipped with a microscopic theory of chirality induced spin effects in electron transfer reactions, it may be possible to answer such questions.

\section*{Acknowledgements}

We would like to thank Anthony Poggioli for useful comments on this manuscript. We are also thankful to Peter Hore for pointing out the connection between this work and the CIDEP effect in the EPR literature. T.P.F and D.T.L. were supported by the U.S. Department of Energy, Office of Science, Basic Energy Sciences, CPIMS Program Early Career Research Program under Award No. DE-FOA0002019.

\section*{Supporting Information}

Further details of the HEOM and QME calculations, a brief analysis of the effects of competing back reactions and singlet-triplet dephasing on spin polarization, and a discussion of the connection between SOC mediated CISS and chemically induced dynamic electron polarization (CIDEP).

\bibliography{bibliography.bib}

\end{document}